\documentclass[12pt]{iopart}

\usepackage{graphicx}
\usepackage{hyperref}
\usepackage{lipsum}
\usepackage{multirow}

\begin{document}

\title[Stellector night sky explorer]{Stellector: a laser aided night sky explorer for teaching astronomy}

\author{M N S Silva, F A Pedersen and J T Carvalho-Neto}

\address{Departamento de Ci\^{e}ncias da Natureza, Matem\'{a}tica e Educa\c{c}\~{a}o, Universidade Federal de S\~{a}o Carlos, Caixa Postal 153, 13600-970, Araras, SP, Brasil}
\ead{jteles@ufscar.br}
\vspace{10pt}
\begin{indented}
\item[]November 2021
\end{indented}

\begin{abstract}

We present a device, created by us and named Stellector, composed by a laser pointer which is precisely guided by two step motors with the purpose to explore and teach astronomy concepts having the real night sky in the background. The electronic part is made of low cost items and the mechanical part is 3D printed. The controller software was written in HTML/Javascript language in order to run in any portable communication device, such as smartphones and tablets. Communication with the Stellector hardware is via Bluetooth standard. These characteristics ensure the necessary portability and autonomy for outdoor astronomy teaching activities. In this work, we sketch the Stellector design and its mode of operation. We also illustrate some teaching activities involving basic night sky observations and astronomy concepts. Finally, we discuss the device limitations, its accuracy and further improvements.

\end{abstract}

\ioptwocol

\section{Introduction}

The systematic observation of the night sky is a very pleasant activity where the regularities and nature of the astronomical objects reveal themselves as beautiful as the star patches that spread throughout the Milk Way. The teaching of basic concepts of observational astronomy can take advantage of these perceptions using the real night sky as a live blackboard \cite{barclay2003}. Due to its intrinsic outdoor nature that demands open skies far away from tall buildings and highly illuminated areas, these activities are mainly employed in non formal educational environments \cite{menezes2018}.

There are many works devoted to explore astronomy teaching activities by the night sky observation \cite{barclay2003, alencar2009}. With the use of basic optical instruments, such as small size telescopes and binoculars \cite{moore2000}, one can observe the hidden features that bring the nature of the celestial bodies closer to that of our own Earth, with their structures, imperfections, differences and similarities, as first appreciated by Galileo Galilei \cite{gingerich2011}.

But is by observing the sky by the naked eye that one can follow the apparent motions of the celestial sphere as a whole along the night, the year and the seasons, the distribution of stars in constellations and asterisms with their associated cultural meanings \cite{alencar2009, spinelli2019, fatima2021}, the characteristic paths followed by the main solar system bodies, the relations with local geographical coordinates, among other characteristics. Sky charts \cite{justiniano2016}, planispheres and celestial globes \cite{ruangsuwan2009, trogello2015} are the main tools useful to guide observers throughout the night sky observations. More recently, we find sky map softwares \cite{hughes2008, zhang2014, morris2018, zotti2021} and smartphone apps \cite{tian2014} that use their internal sensors to centralize the sky map in the direction the device is pointing to.

Specially useful are the laser pointers that can be used to point to the celestial objects in the night sky \cite{bara2010}. With a sufficiently dark environment, the light path of the laser pointer is made visible due its dispersion by particles suspended in the atmosphere and due to Rayleigh scattering. Due to its long reach, people a couple of meters away from the laser pointer are weakly affected by the parallax near the end of the visible light beam. Therefore, people have the perception that the laser beam is touching the celestial dome at the star chosen by the laser pointer operator. It is worth noting that, unlike the proposal of this work, lasers have been used in large research telescopes to compensate for deleterious atmospheric effects through so-called active and adaptive optics techniques \cite{hubin1993}.

Therefore, to take advantage of the light path impinged by the laser, we attached the laser pointer to a pair of step motors (sometimes they are simply called steppers), which can move the laser beam to specific directions commanded by a software. Such control opens the possibility to a variety of operations and learning activities, such as: (i) celestial bodies identification, (ii) ray tracing of the constellation and asterism lines for different cultures \cite{alencar2009}, (iii) ray tracing of the constellation borders, (iv) ray tracing of the Milk Way galactic disk, (v) ray tracing of the planets movement along the year relatively to the stars, (vi) identification of the ecliptic and solar positions along the year, (vii) in combination with binoculars or telescopes helping to find faint objects such as nebulae, galaxies, Uranus and main belt objects.

The laser aided night sky explorer thus developed by us was named \textit{Stellector} (a compressed form for Stellar Projector) and it intends to project the dynamics and representations of the sky, using the real open sky itself as a background for its tracings. The main purpose of this work is to describe how it works and how it can be helpful in the teaching of astronomy.

\section{System description}

\begin{figure}
	\caption{\label{fig_4} Stellector schematic diagram. The dashed line delimits the Horus apparatus.}
	\centering
    \includegraphics[width=1.0\columnwidth]{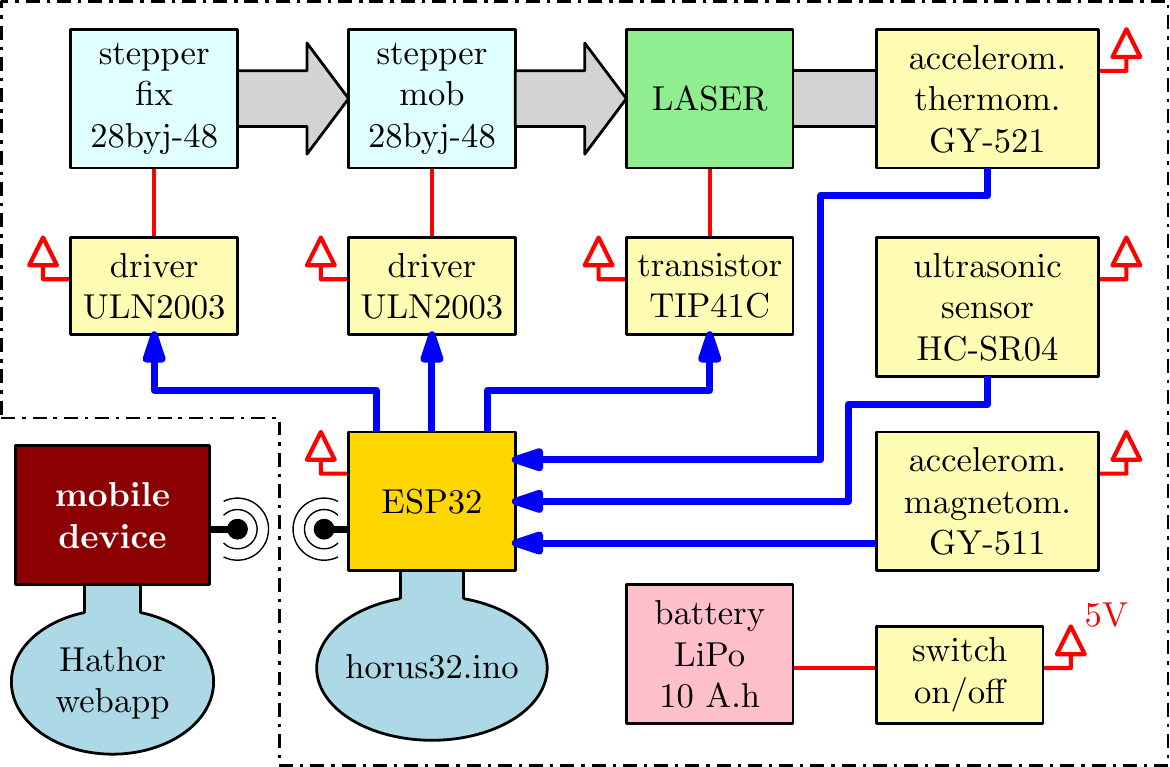}
\end{figure}

Figure~\ref{fig_4} depicts a schematic diagram in which the various parts of the Stellector are shown along with their interrelationships. Rectangular shapes represent hardware elements (described in section~\ref{ssec:hardware}) and oval shapes represent the software that is described in section~\ref{sec:software}. In light blue are indicated the step motors, in green the laser and in light yellow the sensors or accessory electronic circuit boards. The connections in gray indicate mechanical connections between components, where the direction of the arrow implies that the next element is moved by the previous one, otherwise the connection is fixed. Blue connections indicate the direction of digital communication between components and the microcontroller. The connections in red indicate the power supply, where all the red arrows are connected to the same power source at the 5~V output of the on/off switch that connects the battery to all the hardware elements.

\subsection{\label{ssec:hardware} Hardware}

\begin{figure}
	\caption{\label{fig_1} Stellector general arrangement photographs. (a) Tripod mount as expected for sky observations. (b) Close view of the Horus hardware next to a smartphone for equipment operation.}
	\centering
    \includegraphics[width=1.0\columnwidth]{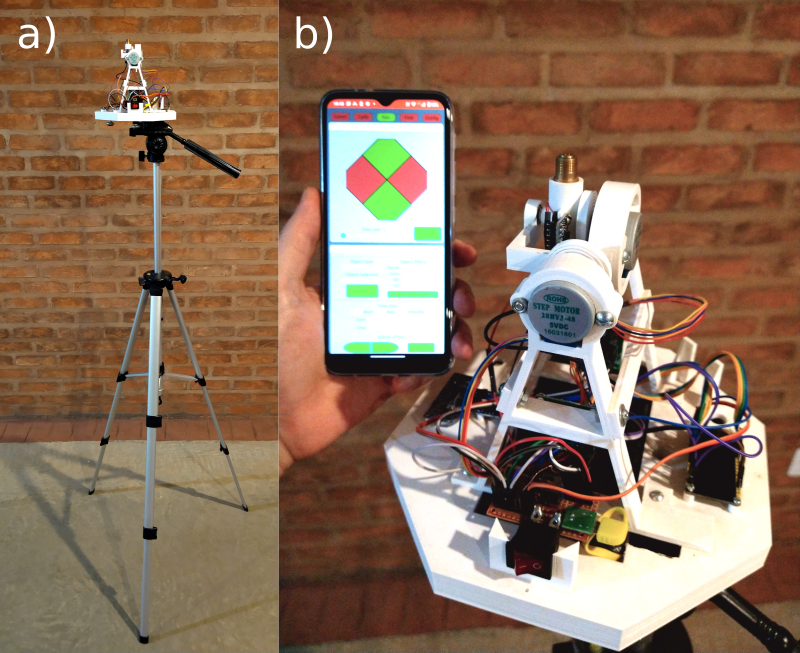}
\end{figure}

The hardware named \textit{Horus}, after the Egyptian god of the sky, is composed of a mechanical and electronic part. It is showed in figure~\ref{fig_1}(a) mounted on a tripod. In figure~\ref{fig_1}(b) we highlight the Horus, which has 21.5 cm in width and approximately 15 cm in height. There is no need to be wired to an external power source as it has built-in rechargeable battery. All operation and communication with Horus is done by a portable device via Bluetooth standard, such as an ordinary smartphone as seen in the upper left corner of figure~\ref{fig_1}(b). These features ensure the autonomy and versatility necessary for outdoor use.

\subsubsection{\label{sssec:mechanics} Structure }

\begin{figure}
	\caption{\label{fig_2} Perspective view of Horus pieces consisting of octagonal base (in purple), two support towers (in yellow and cyan), mobile step motor support (in red and green), and laser support (in dark blue). Drawing developed with FreeCAD software.}
	\centering
    \includegraphics[width=1.0\columnwidth]{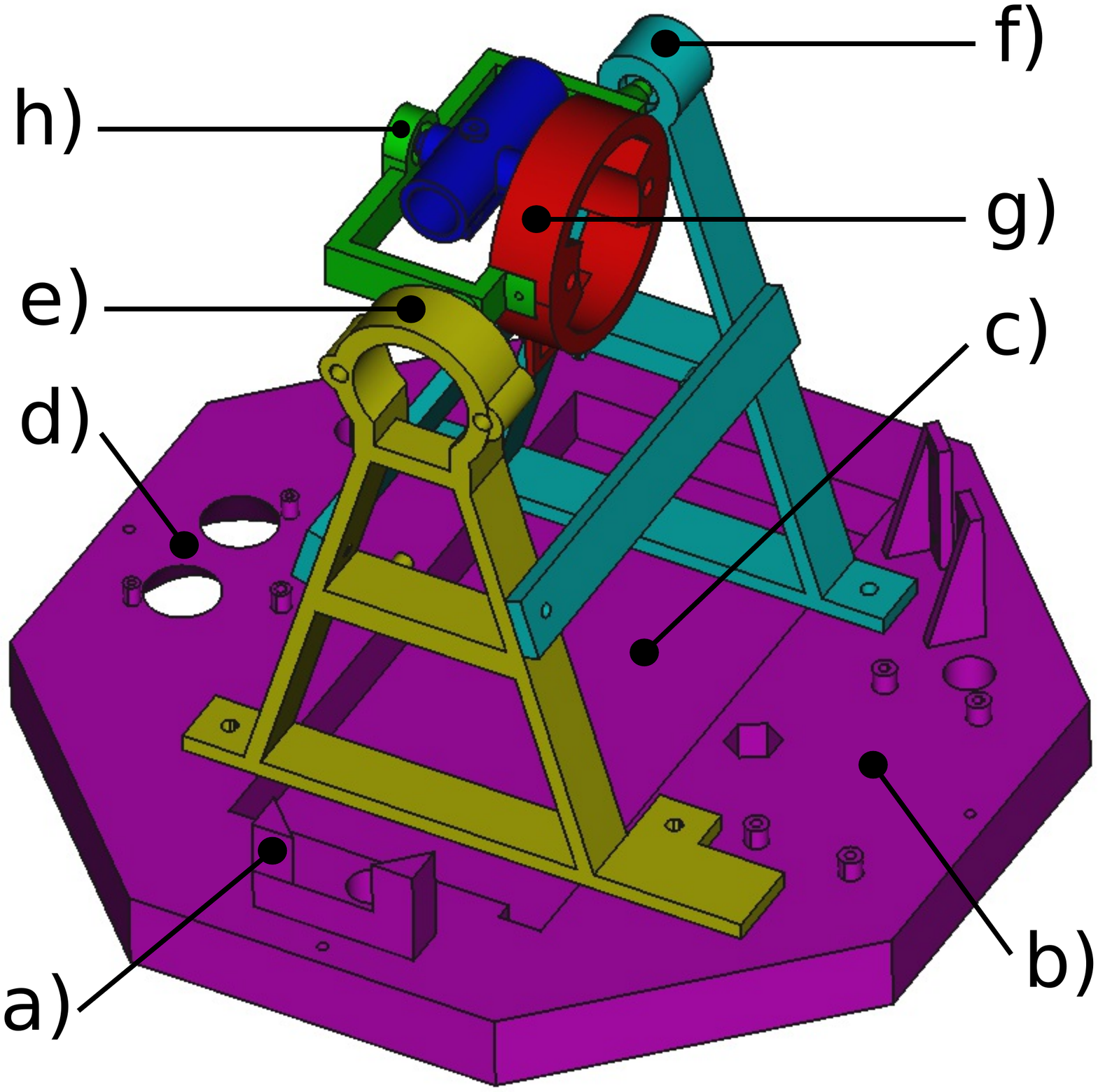}
\end{figure}

The mechanical structure of Horus consists of a flat octagonal base, two triangular-shaped support towers and two mobile supports with rotation axes perpendicular to each other. Figure~\ref{fig_2} contains the perspective view of the structure drawn in CAD (Computer-aided design) software. Each colored piece was 3D printed separately and later fixed with common metal screws.

We now describe the items shown in figure~\ref{fig_2}. Item a) corresponds to the on/off switch fitting of the electrical part. The electrical energy distribution circuit for all other electronic, electromechanical and laser components is fixed (welded) to this same switch. Item b) corresponds to the location of the microcontroller board (we used the ESP32 board discussed in section~\ref{sssec:electronics}) responsible for the logical-computational interface between step motors, sensors and laser with the user's control device (smartphone, tablet, laptop, desktop, etc) via Bluetooth communication. Item c) corresponds to the rectangular cutout for fitting the rechargeable battery. We choose a lithium ion polymer model with a capacity of 10 A.h, which is used to recharge smartphones, known as power banks. The main advantage of the power banks is that the electronics responsible for controlling the charge and discharge of the battery are already incorporated into the set. The 5V power bank output is connected to the power circuit that is part of item a). From our estimates of the equipment's energy consumption, at full load, the battery is capable of powering the Horus components for at least 6 hours, which is enough for a full night observation. Item d) corresponds to the cutouts and holes for fixing the ultrasonic sensor and the level accelerometer for measuring the distance from the Horus base to the ground. This is a security feature that is discussed in section~\ref{sssec:electronics}. The triangular towers support the step motors, the laser and laser-coupled accelerometer assemblies. The functions of these items are also explained in section~\ref{sssec:electronics}. Step motor drivers are screwed into mounting holes located in the crossbars of the towers. Item e) corresponds to the fixing ring of the fixed step motor, responsible for the variations in the azimuthal angular coordinate of the laser. Item f) corresponds to the fitting of the step motor support shaft via a 9.5 mm outer diameter bearing. Item g) corresponds to the fitting of the mobile step motor, responsible for the variation in the laser's polar angular coordinate. Item h) corresponds to the fitting of the laser support shaft also via a 9.5 mm outer diameter bearing.

\begin{figure}
	\caption{\label{fig_3} Perspective view of the Horus laser support. Drawing developed with FreeCAD software.}
	\centering
    \includegraphics[width=1.0\columnwidth]{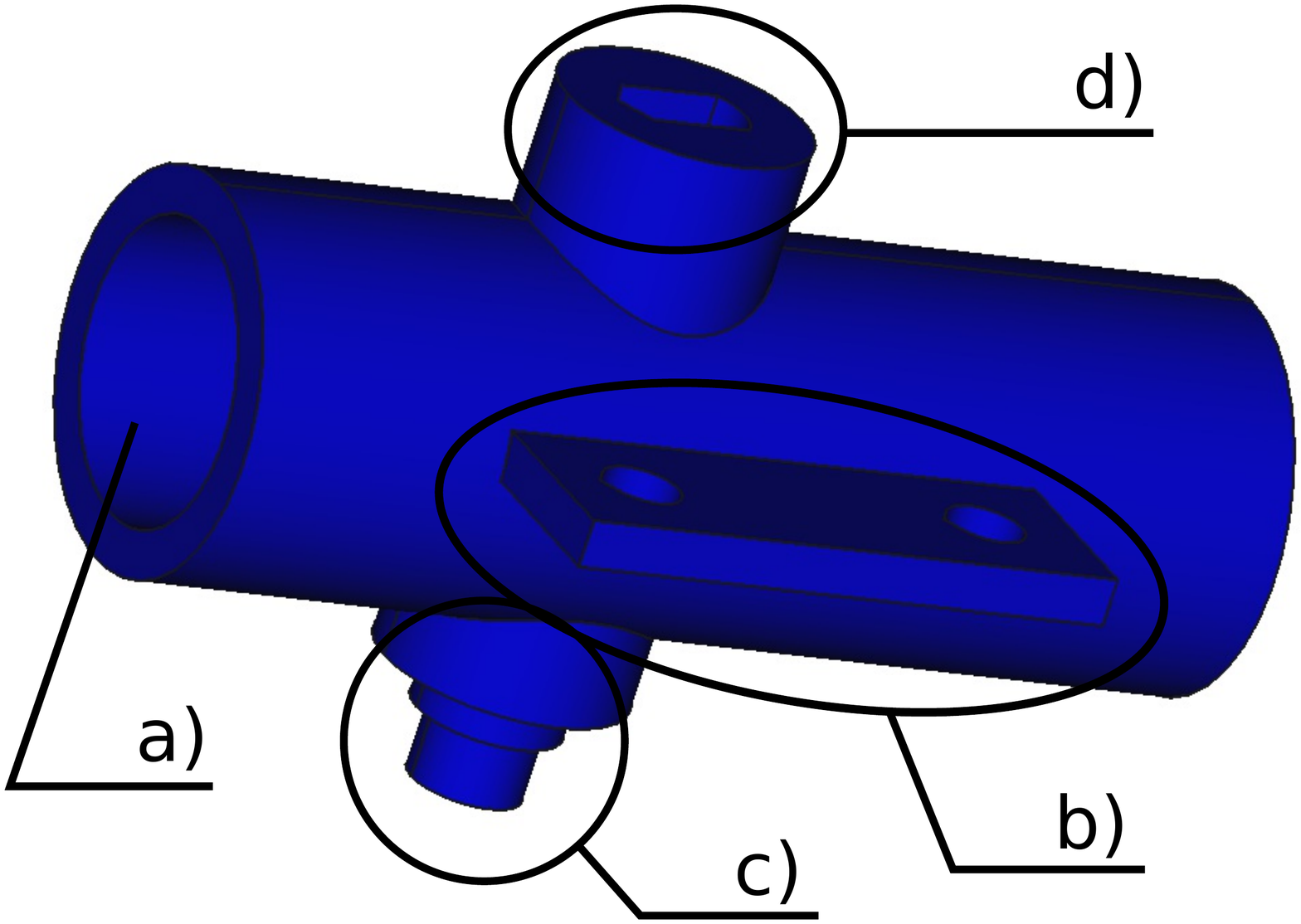}
\end{figure}

Figure~\ref{fig_3} contains the drawing of the laser support separately, in which item a) corresponds to the fitting of the laser cylindrical housing and item b) corresponds to the base and fixing holes of the laser orientation accelerometer. The purpose of this accelerometer is to provide a reference direction which is explained in more detail in section~\ref{sssec:electronics}. Item c) corresponds to the laser support shaft and item d) corresponds to the mobile motor shaft fitting.

\subsubsection{\label{sssec:electronics} Electronics and electromechanics}

The main component of the Horus electronics is the development board based on the ESP-WROOM-32 module. It consists of a microcontroller connected to several basic peripherals integrated for its operation (Wi-Fi and Bluetooth devices, USB serial communication, GPIO pins, among others). The use of this system is very similar to the Arduino platform, which has been applied in several projects, such as industrial/residential automation, art and entertainment, teaching and science in general.

The ESP32 outputs used in Horus consist of 8 logic level pins for the step motors (4 pins for each motor) and 1 logic level pin for the laser drive. Two other pins are used for I2C communication with the accelerometers and ultrasonic sensors.

We used the step motor model 28byj-48, which has 32 steps per complete revolution and is one of the most used step motors in Arduino projects due to its great cost/benefit relation. This motor has an internal gear set with a reduction ratio of 64:1. So, effectively, each step of the motor corresponds to an angular displacement of $\frac{360^{\circ}}{32\times64}\approx0,18^{\circ}$. Consequently, this is the angular accuracy in spherical coordinates. By way of comparison, this value corresponds to a little more than a third of the angular size of the full moon as seen from Earth, a satisfactory accuracy for naked eye observations.

The step motors were configured in a similar way to an equatorial mount, where the stepper responsible for the variation of the right ascension coordinate (azimuthal angle in spherical coordinates) was named as fixed motor. The stepper responsible for the declination coordinate (polar angle in spherical coordinates) was named as mobile motor. However, the fixed motor shaft is not oriented parallel to the Earth's rotation axis as usual, but parallel to the local horizon. This orientation is important to avoid geodesic traces too close to the fixed motor axis, which would require very large and, consequently, time-consuming angular displacements of the mobile motor.

Although the 28byj-48 is a low cost, small size and good angular accuracy step motor, there is a gap of a few degrees in the shaft rotation that causes (when the direction of rotation is reversed) several false steps to be taken. This is a very detrimental feature for the Stellector, and to minimize it, we attached a flat torsion spring -- also 3D printed -- to the shaft of each motor in order to keep the shaft always tensioned, thus eliminating false steps when the steppers rotation direction is reversed.

To establish a reference direction to which the laser is pointed, we attached an accelerometer (model GY-521) to the laser mount. It measures the three Cartesian components of the acceleration vector to which it is subjected. At rest, it therefore measures the components of the gravitational acceleration. From the measurement of these components, the laser is automatically pointed to the zenith, at the beginning of the Horus' operation, through a programming routine written in the horus32.ino code \cite{stellector2021} saved on the ESP32 board.

The laser device, in turn, was extracted from a laser pointer with wavelength of 532 nm (green color). The laser device has a cylindrical metal casing and was fitted to the laser support indicated in figure \ref{fig_3}(a). For its driver we have elaborated a simple switching circuit using a TIP41C power transistor. By connecting one of the ESP32's digital outputs to the transistor's base and applying a high logic level of 3.3 V to the base, our circuit results in a voltage of approximately 2.5 V to the laser. With this voltage we can already observe a significantly pronounced trail in the night sky in places with low light pollution. In the original laser pointer case, approximately 3 V is applied from two alkaline batteries. We prefer to work with a lower voltage to extend the laser lifetime, since astronomical exposures with the Stellector can be long. Still, in places with excessive lighting it is possible to increase this power by adjusting a potentiometer connected in series to the laser device.

Reference \cite{stellector2021} contains the electronic diagram of the circuit connected to the Horus on/off switch. It shows the laser driver with the TIP41C transistor, as well all connections among the electronic sensors, the microcontroller and the electromechanical devices of the Horus.

One of the main safety concerns in using the Stellector is the risk of the laser light hitting the eyes of people around Horus, including the person operating it \cite{bara2010, mclin2011}. To minimize this risk, we included an ultrasonic sensor, model HC-SR04, in combination with an accelerometer in the octagonal base (figure \ref{fig_2}(d)) in order to measure the apparatus distance to the ground. We have devised a reasonably safe way to use the Stellector by securing it to a tripod at least 1.70 m above the ground. By adding the height of the fixation towers (approximately 15 cm), we guarantee a laser height of at least 1.85 m from the ground, minimizing the risk of the laser reaching people's eyes. Height measurement is done by sending an ultrasound pulse towards the ground and measuring its echo time. If the measured height is less than 1.70 m, the code on the ESP32 board (horus32.ino) does not allow the laser to operate. The accelerometer used in conjunction with the ultrasonic sensor has the function of ensuring that the projector base is horizontal so that the ground clearance is as vertical as possible. The model GY-511 (LSM303DLHC) was used for the level accelerometer, which also includes an electronic compass. With this, the GY-511 can additionally be used to assist in determining the initial alignment of the Stellector, making it easier for the lay user to identify the calibration stars (procedure discussed in section \ref{ssec:operation}). The two sensors, HC-SR04 and GY-511, are low cost and widely used in Arduino projects.

All Horus 3D printed pieces were designed with FreeCAD software via Python scripting code. These codes and the generated model files are available in reference \cite{stellector2021}.

\subsection{\label{sec:software} Software}

There are two distinct programming platforms used in the Stellector. The first one is used to produce the operating and communication instructions for the ESP32 board housed in the Horus enclosure. We employed the same programming language used on the Arduino boards, which is basically the C/C++ language plus some specific functions for hardware control and communication. The corresponding programming code was called horus32.ino and is shown in figure \ref{fig_4}.

The second platform consists of the one used to build the user interface, establish the two-way communication with Horus and perform the more advanced and, consequently, more processing- and memory-heavy computational tasks. We used Javascript/HTML languages to develop the web application called Hathor, in reference to the Egyptian goddess consort of the god Horus\footnote{The ancient Egyptian deities have a much richer interpretation and understanding than just the "God of Sky" and his consort. We limit ourselves to a very superficial meaning for labeling purposes only.}. JavaScript is a high-level language, with an extensive documentation and programming community, and with a wide range of specific libraries. But most importantly, web applications developed in JavaScript do not need to be installed on the device. They are interpreted by the web browsers themselves and thus become more universal, where the same code runs equally well on any platform, from Windows, macOS, or Linux desktops and laptops to Android or iOS tablets and smartphones. We found four JavaScript libraries that provided important functions and data for this project. The first one, called THREE.js, is responsible in this project for performing many vector operations and coordinate transformations between the rectangular and spherical systems. The second -- d3-celestial (a star map with d3.js) -- and the third -- orb.js (JavaScript Library for Astronomical Calculations) -- are used to obtain the equatorial coordinates of stars and objects in the Solar System, lines and boundaries of constellations and various asterisms. The fourth library -- fmin (Unconstrained function minimization in Javascript) -- is used for minimizing the function that calculates the orientation of the local spherical coordinate system (defined by the orientation of the Horus) relative to the astronomical equatorial coordinate system. Besides those ready-made libraries that were very useful, we developed JavaScript code with specific functions for this project \cite{stellector2021}.

\subsection{\label{ssec:operation} Mode of operation}

\subsubsection{\label{sssec:setup} Initial setup}

First, one should check that the Horus battery is fully charged. This condition is indicated by LEDs on the body of the power bank containing the battery. One should look for a dark place with few or no buildings around -- not only because the horizon is less obstructed, but also to avoid the risk of the laser beam hitting the eyes of distant people located above the level of the laser pointer. Having determined the observation site, one should look for a position where the ground is as horizontal as possible so that the tripod can be mounted and set to its maximum height. After attaching the Horus to the tripod, one can just flip the power switch and access the Hathor web app on the mobile device -- preferably a smartphone -- which should be at its minimum brightness setting so as not to dazzle viewers' eyes.

The Hathor web application consists of five windows, where only the \textit{Find} window has not yet been implemented. Each window is made up of several tabs that can be pushed up or down by tapping its title. Below we explain their main features, but we show figures of just a few of them to fit the space of this article.

\subsubsection{\label{sssec:communication} Communication}

The first step is to establish communication with Horus, which is done through the communication window -- abbreviated by \textit{Comm}. When one clicks on the \textit{Start or restart communication} button, the communication via Bluetooth protocol starts (a web browser compatible with the Web Bluetooth API must be used). In this window there is also a log text field, in which all commands sent to Horus and all information read from it are notified. It has a more technical function, useful for users interested in more detailed operation of the equipment and for controlling any software failures. The Comm window is only used once when the equipment is started. Occasionally, there may be a loss of communication between Horus and Hathor, making it necessary to re-establish communication via the \textit{Start/restart comm} button.

\subsubsection{\label{sssec:calibration} Calibration}

The next step involves defining the Horus orientation in relation to the equatorial coordinate system, a procedure we call calibration and which is abbreviated by \textit{Calib} in the Hathor app.

In principle, this initial calibration could be done knowing the local geographic coordinates -- latitude and longitude -- and the geographic north-south orientation. The first is easily obtained over the internet or via GPS from the smartphone itself with quite satisfactory accuracy. However, the second is more difficult to be obtained with precision, being the most ready method the orientation by the stars (notably by the Southern Cross constellation in the southern hemisphere and by the Polaris star in the northern hemisphere).

We prefer to use as a calibration procedure the measurement of the local coordinates of some known visible stars using the step values of the step motors, which provide the spherical angular coordinates accurate to 0.18°.

\begin{figure}
	\caption{\label{fig_5} Controller tab of the Hathor web app available at the \textit{Nav}, \textit{Calib} and \textit{Find} windows.}
	\centering
    \includegraphics[width=0.95\columnwidth]{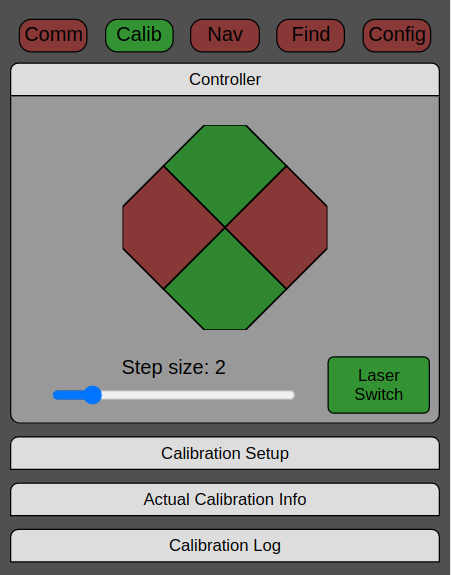}
\end{figure}

This window has four tabs that we describe next. The figure \ref{fig_5} shows the calibration window of the Hathor app with just the Controller tab opened.

\begin{itemize}
    \item \textit{Controller tab.} This is the manual control for laser movement. It has 4 directional buttons: fixed motor in the right and left hand directions, and mobile motor in the up and down directions. Below the directional buttons there is a slider for the size of the angular displacement of the motor shaft at each click of the directional buttons, which can take on values of 1, 2, 4, 8, 16, 32, 64 or 128 steps. Remembering that each step of the motor is approximately 0.18º. Finally, to the right of this control there is a button for turning the laser light on or off. Due to its importance, this \textit{Controller} tab is also available in other windows.
    
    \item \textit{Object Selection in Calibration Setup tab.} It contains four fields to search for the desired calibration star: (i) \textit{Name}, to search for the star's western proper name, (ii) \textit{Cons}, for search for the western name of the constellation to which the star belongs, (iii) \textit{HIP} for the Hipparchus code of the star and (iv) \textit{HD} for the HD code of the star. In the case of selection by Solar System objects (basically the Moon and the planets), only the field \textit{Name} provides identification of the object. When typing the first 3 characters of any of the four fields, the application filters all stars that fit these characters and displays them in the selection box identified by \textit{Object selected}. Finally, this tab has the \textit{ADD CALIB OBJECT} button, which must be activated when the laser is pointing at the chosen star in the \textit{Object selected} selection box. By clicking on this button, the selected star is added to the \textit{Calibration List} field discussed below.
    
    \item \textit{Calibration List in Calibration Setup tab.} Contains the list of celestial objects already chosen for calibration. To the right there are buttons \textit{remove this} for removing a particular star from the calibration list and \textit{remove all} for removing the entire list. After the inclusion of two or more stars in this list, it is possible to calculate the calibration of the orientation of the Horus local coordinate system, which is done by clicking on \textit{CALC} button. The result of this calibration is displayed in the \textit{Calibration Log} tab. If the user considers that the calibration was successful, he clicks on \textit{ACCEPT} button, making any navigation command from then on take into account the orientation obtained from this calibration.
    
    \item \textit{Actual Calibration Info tab.} It displays a text field with the information of the stars used in the current calibration obtained by pressing the \textit{ACCEPT} button.
    
    \item \textit{Calibration Log tab.} It displays a text field with information on all calculated calibrations, including those that were not chosen by the \textit{ACCEPT} button.
\end{itemize}

\subsubsection{\label{sssec:navigation} Navigation}

After calibration using the above procedures, the Stellector is ready to be used in exploring the night sky. And for that, the navigation window -- abbreviated to \textit{Nav} -- is the main feature.

This window contains three tabs, besides the Controller one, that we describe next.

\begin{figure}
	\caption{\label{fig_6} \textit{Sky Objects} tab of the Hathor web app available at the \textit{Nav} window.}
	\centering
    \includegraphics[width=0.95\columnwidth]{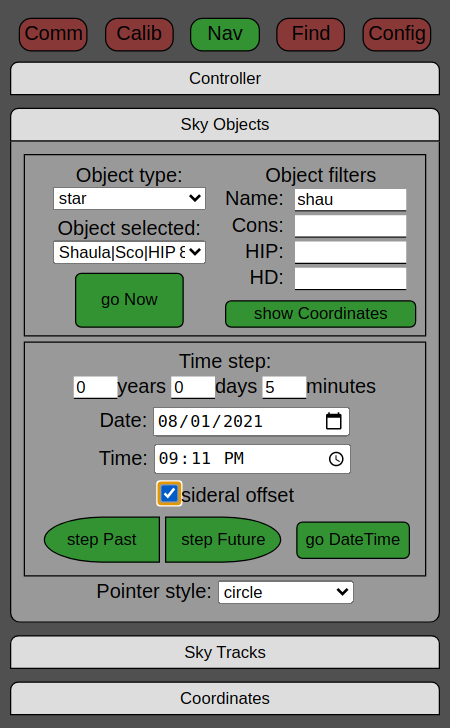}
\end{figure}

\begin{figure}
	\caption{\label{fig_7} \textit{Sky Tracks} tab of the Hathor web app available at the \textit{Nav} window.}
	\centering
    \includegraphics[width=0.95\columnwidth]{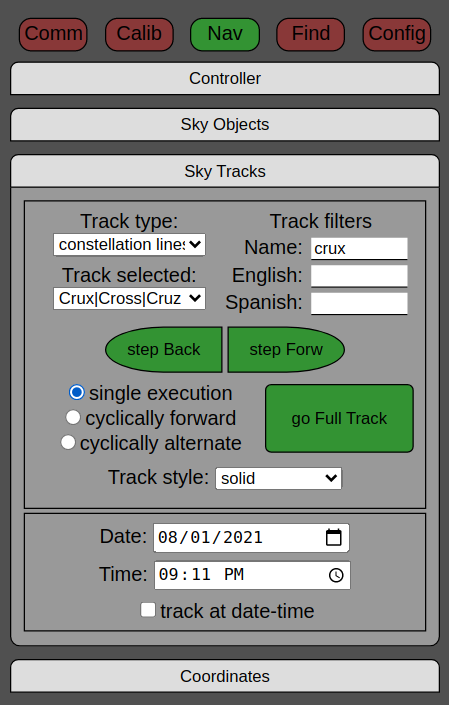}
\end{figure}

\begin{itemize}
    \item \textit{Sky Objects.} This tab is shown in figure \ref{fig_6}. With it, one can point to individual celestial objects. The \textit{Object Type} options are: (i) solar system (with equatorial coordinates calculated by the orb.js library), (ii) star, (iii) deep sky and cluster, (iv) messier and (v) constellation center (with equatorial coordinates obtained from the d3-celestial javascript library). Then, the object is selected by typing the first letters in the \textit{Object filter} field, similarly to what was explained in the calibration window. The selection box \textit{Pointer style} allows one to choose the option \textit{point} which corresponds to the laser to remain fixed exactly on the coordinate of the star and the option \textit{circle} which corresponds to the laser to execute a circular movement around of the star. By pressing the \textit{go Now} button, the laser is pointed at the currently selected star at the current moment. By pressing the \textit{go DateTime} button, the laser is pointed at the object at the chosen \textit{Date} and \textit{Time}. In addition, it is possible to temporally advance and rewind the position of the object through a pre-chosen \textit{Time step} with the \textit{step Past} and \textit{step Future} buttons. When taking the steps, it is possible to disregard sidereal movement by checking the \textit{sidereal offset} option. This is an especially useful option for showing the movement of solar system objects relative to the fixed stars. Examples of using this functionality are explored in section \ref{sec:activities}.
    
    \item \textit{Sky Tracks.} This tab is shown in figure \ref{fig_7}. The purpose of this tab is to navigate and trace lines and boundaries involving the constellations, asterisms and the ecliptic. With it, it is possible to make the cyclical tracing, or one step at a time, among the stars of the chosen constellation or asterism, allowing a wide visualization of the sky region and the set of stars covered by the constellation. Tracings can be taken at the current moment or at the positions occupied by the stars in the chosen Date and Time.
    
    \item \textit{Coordinates.} In this tab it is possible to check or access the coordinates directly, without the need for them to be associated with the position of a particular celestial object. The units of the spherical coordinates indicated for the laser can be in \textit{steps}, which correspond to the number of steps of the fixed (Fix) and mobile (Mob) motor, or \textit{equatorial}, which correspond to the equatorial coordinates RA (Right Ascension) and Dec (Declination). The \textit{go Equatorial} and \textit{go Steps} buttons guide the laser to the chosen equatorial coordinate or step motor coordinate, respectively. The \textit{read Coords} button reads the coordinates in which the laser is positioned and the \textit{go Zenith} button automatically takes the laser to the zenith direction using the reading of the accelerometer fixed to the laser as explained in section \ref{sssec:electronics}.
\end{itemize}

We are also developing a \textit{Custom path} tab. For this tab, the user could load or type in a sequence of coordinates that correspond to a specific laser trace. With this, we intend to use predetermined tracings that could indicate asterisms or constellations of different cultures, a pre-planned journey by the exhibitor, etc. For example: one could indicate the procedure to find the south celestial pole using the Southern Cross constellation. Activities of this type are explored in the section \ref{sec:activities}.

\subsubsection{\label{sssec:configuration} Configuration}

This window contains settings that allow for more specific and advanced customizations.

\begin{itemize}

\item \textit{Pointer Config.} In this tab, one can choose the laser linear and circular tracing speed, the circular aperture angular radius, among other tracing parameters.

\item \textit{Speech Config.} Here, one can choose to have the Hathor app pronounce the name and designations of celestial objects as they are traversed by the laser using the \textit{step Back} and \textit{step Forw} buttons on the \textit{Sky Tracks} tab of the \textit{Nav} window.

\item \textit{Calibration Config.} In this tab, additional parameters to be optimized in the calibration of the Horus coordinate system are considered. The \textit{fix} and \textit{mob angle stretching} parameters allow one to apply a multiplicative factor to the angular size of each step of the step motors. This may be necessary due to the tension produced by the torsion springs used to avoid the missteps discussed in section \ref{sssec:electronics}. The \textit{laser} and \textit{mob axis tilt} parameters allow for a correction in the angle between the laser direction and the mobile motor axis and between the mobile motor axis and the fixed motor shaft. Ideally, both angles are 90°, but constructive failures and stresses in the supports can produce small deviations from that.

\end{itemize}

\section{\label{sec:activities} Activities}

To illustrate the possible didactic uses of the Stellector, we propose four activities that explore basic concepts of observational astronomy.

Calibration of the Stellector must be done in advance so that the system knows how to determine the orientation of the Horus local coordinate system with respect to the celestial equatorial coordinate system. The procedure is similar to that adopted in the GoTo systems of motorized telescopes. The user identifies in the sky the names or designations of a set of visible stars not very close to each other and points the laser (telescope in the case of GoTo systems) at these stars, informing the system the identification of the star being pointed out. Sky charts or sky viewing apps can be used to help with this identification. Detailed commands for calibration on the Stellector were described in section \ref{ssec:operation}. At least two stars are required for the unambiguous determination of the relative orientation between coordinate systems. However, due to structural imperfections in the mechanical part of Horus, additional parameters must be optimized, implying the need for a greater number of stars. Suggested numbers based on our experimentation are discussed in section \ref{sec:experimental}.


\subsection{\label{ssec:celestialsphere} The celestial sphere and geolocation}

The purpose of this activity is to show the apparent movement of the stars due to the Earth's rotation around its axis, establishing relationships with the temporal concepts of day length and spatial geographic coordinates.

\subsubsection{\label{sssec:fixedStars} Movement of fixed stars.} In the Hathor app, select a visible star close to the south or north celestial pole (depending on the location) in the \textit{Sky Object} tab of the \textit{Nav} window. Click on \textit{go Now} to show its current position to the public. Choose a \textit{Time Step} of a few minutes and press the \textit{step Past} or \textit{step Future} button repeatedly. With this, the laser will show the position of the star over time, showing the circular trajectory concentric to the visible celestial pole at the location. When executing these commands, the exhibitor can mention the size of the temporal step so that people can quantify the relationship between the star's movement and the passage of time. This procedure can be repeated for stars with other declinations, including those near the celestial equator. In this case, it is also possible to take a short future time step with the \textit{step Future} button -- 3 minutes for example -- while the presenter does some theoretical explanation about astronomical concepts. After this interval, everyone will be able to verify if the star has reached the laser position.

\subsubsection{\label{sssec:celestialCircles} Celestial circles and cardinal points.} In the Hathor app, in the \textit{Coordinates} tab of the \textit{Nav} window, type +90° in the \textit{Dec} field for the northern hemisphere or -90° for the southern hemisphere and press the \textit{go Equatorial} button with the \textit{Pointer Style} in the \textit{point} option. This will indicate the chosen celestial pole, also indicating that the Earth's axis of rotation is parallel to the laser beam. Then, on the \textit{Pointer Config} tab of the \textit{Config} window, choose the value of the \textit{Circle aperture} parameter with the angle that will correspond to the geographic circle you want to highlight. For example, to trace with the laser the celestial polar circle, tropic and equator, choose, respectively, 23.5°, 66.5° and 90° for the parameter \textit{Circle aperture}. Then, go back to the \textit{Nav} window and press the button \textit{go Equatorial}, now with the \textit{Pointer Style} in the \textit{circle} option. The laser will execute the chosen circular trace indefinitely until another command is executed.

\subsubsection{\label{sssec:locations} Simulation of other locations.} It is possible to trace the celestial circles and the Earth's rotation axis as seen at a different target geographic latitude. To do this, just carry out the previous activity replacing the declination coordinate in the \textit{Dec} field with $(\theta_{t}-\theta_{l})$, where $\theta_t$ and $\theta_l$ are the target and local latitudes, respectively.

In all procedures of this activity, the \textit{sidereal offset} option must not be checked, so that the movement of the celestial sphere is not compensated.

\subsection{\label{ssec:ecliptic} The ecliptic and the seasons}

The objective of this activity is to show the apparent movement of the Sun throughout the year and, with this, to explore the concept of the seasons. We think of two ways to achieve this goal with the Stellector.

\subsubsection{Position of the Sun throughout the year in relation to the stars.} In the Hathor app, select the \textit{Sun} in the \textit{Sky Object} tab of the \textit{Nav} window. As the Stellector obviously only works at night, you should choose a \textit{Date} approximately six months before or after the current date. Then, choose a \textit{Time step} of a few days and press the \textit{step Past} and \textit{step Future} buttons repeatedly to show the audience the positions that the Sun travels throughout the year relative to the fixed stars. In this case, the \textit{sidereal offset} option must be checked so that the motion of the fixed stars is disregarded. Interspersed with this exposure, the tracings of the celestial tropics of Cancer and Capricorn can be performed following the procedure in section \ref{sssec:celestialCircles} with the purpose to discuss the relationship of the tropics with the positions of the Sun and its effects for the seasons.

\subsubsection{Drawing of the ecliptic.} Alternatively, it is possible to show the path that the Sun takes in the sky in relation to the stars by selecting the \textit{ecliptic} option under \textit{Track type} in the \textit{Sky Tracks} tab of the \textit{Nav} window. By clicking the \textit{go Full Track} button with one of the options \textit{cyclically} checked, the laser will repetitively trace the path of the ecliptic until some other command is executed. With this, the exhibitor will be able to explain the astronomical concepts without having to constantly touch on the Hathor app.

\subsection{Constellations and asterims}

The objective of this activity is to show how the sky is divided into different constellations and how we can identify the main stars of the constellations through the geometric patterns they form and their relationship with geographic coordinates.

\subsubsection{Lines of the constellations.} Select the option \textit{constellation lines} in \textit{Track type} in the \textit{Sky Tracks} tab of the \textit{Nav} window and then select the desired constellation with the help of the \textit{Track filter}. By clicking \textit{go Full Track}, the laser traces lines connecting the main stars of the constellation to show the pattern of the western shapes (e.g. Scorpius, Orion, Ursa Major, Southern Cross, etc.). There are several \textit{Track Style} options that involve: (i) \textit{solid} -- continuous and direct tracing from one star to another, (ii) \textit{circle} -- tracing circles around each star in the constellation, among other combinations of tracing style. The \textit{single execution} option implies that the trace is executed only once for that set of stars in the constellation. The \textit{cyclically forward} option implies the repetitive tracing in which, upon reaching the last star of the constellation, the laser returns to the first one and repeats the tracing until another command is given in the Hathor app. The \textit{cyclically alternate} option also performs repetitive tracing, however, when reaching the last star, the tracing direction is reversed. In the \textit{Pointer Config} tab of the \textit{Config} window, it is possible to choose the laser's scanning speed among other tracing parameters. If the option track at \textit{Date-Time} is checked, the constellation trace is executed according to its position in the chosen \textit{Date} and \textit{Time}. This function is useful to show the annual periodicity and the position that the different constellations occupy in the local sky throughout the year. Each time one clicks on \textit{step Back} or \textit{step Forw} buttons, a line is drawn sequentially to each star in the constellation. If some of the options in the \textit{Speech Config} tab of the \textit{Config} window are checked, clicking on those step buttons will cause Hathor to pronounce the names and designations of the star in the selected language.

\subsubsection{Borders of the constellations.} For this, the option \textit{constellation bounds} in \textit{Track type} must be selected in the \textit{Sky Tracks} tab of the \textit{Nav} window. Otherwise, the options are similar to the \textit{Constellation Lines} procedures described above. The difference is that the tracing is not performed between the coordinates of the stars, but between the celestial coordinates that define the boundaries of the constellations. Consequently, the Speech options don't work here.

\subsubsection{Asterisms.} The procedures are identical to those for the \textit{Constellation Lines}, and the \textit{asterisms} option in \textit{Track type} must be selected. Asterisms are alternative forms to constellation lines, useful in identifying other constellations, geographic coordinates, and cardinal points.

\subsection{The planets}

The objective of this activity is to show the differential movement of the planets in the solar system in relation to the fixed stars and the region of the sky occupied by them.

One selects one of the visible planets in the \textit{Sky Objects} tab and indicates its position with the laser at the current time or other dates, much like the procedure described in the Position of the Sun in section~\ref{ssec:ecliptic}. When performing the position sweep of the planet over the months in relation to the fixed stars, it is possible to see the location close to the ecliptic and the retrograde motion that occurs when the direction of motion of the planet changes in relation to the fixed stars. The discussions that can be carried out when exploring this type of activity are extensive.

\section{\label{sec:experimental} Experimental characterization}

The first test concerned the safe operation mode, which is the Stellector default mode. It was observed that if the Horus base was positioned at a height below the established threshold set in the horus32.ino code (1.70 m), the path execution was not executed and the laser was not turned on. At the same time, in the Hathor app a message informing the unsafe height was displayed. The same happened if the Horus base was off leveled by more than 3° with respect to the horizontal plane.

We also tested the equipment in the unsafe operation mode by switching the corresponding key in the Horus hardware. In this case, the laser paths were executed even if it could point to the participants eyes. As a warning of this risk, a constant message in red was displayed in the Hathor app indicating the unsafe mode operation.

The next step was to perform the calibration procedures using some visible stars and planets in order to find the orientation of the local reference frame with respect to the equatorial one. The system has seven parameters to be optimized. Three angles related to the orientation of the Horus local reference frame in the equatorial frame and four parameters related to imperfections of the experimental system, which are (i) stretching of the angular step size of each step motor due to the tension produced by the torsion spring used to eliminate the false steps when reversing the stepper rotation direction, as explained in section~\ref{sssec:electronics}, and (ii) an angular deviation from 90° of the laser direction respective to the mobile step motor axis direction, and the same type of deviation of this latter axis respective to the fixed step motor axis direction.

The calibration was performed on September 20 2021 19:00 GMT-0300 (Brasilia Standard Time), at the geographic coordinates of 22.36°S 47.38°W. We started pointing the laser to Vega (Alpha Lyrae) and to the Jupiter planet. By performing the calibration, by pressing the \textit{CALC} button in the calibration window, the optimization resulted in a perfect fitting for the two celestial bodies position. This was due to the system overdetermination, since the number of unknowns was smaller than the number of parameters. This resulted in a poor calibration, where some stars were missed by the laser by more than 3°. By adding more calibration stars, we started getting better results until a total of five celestial objects, above which the calibration enhancement was negligible. Therefore, the following objects were used for calibration: planets Jupiter and Venus and stars Vega (Alpha Lyrae), Altair (Alpha Aquilae) and Antares (Alpha Scorpii). We tested this calibration by asking the Hathor app to point the laser to the following objects: Saturn planet, Deneb (Alpha Cygni), Atria (Alpha Trianguli Australis) and Spica (Alpha Virginis). Among these four objects and the five calibration objects earlier mentioned the worst result was obtained for Deneb, where the laser missed the star by 0.22°. This deviation corresponds to approximately half the angular size of the full Moon.

Having obtained a good calibration, we tested all functionalities of the Navigation window, including the ray tracing of the constellation lines and borders with various laser pointer styles and the steps future and past for stars and planets. With these steps, it was possible to follow the retrograde motion of the Jupiter planet as it started to reverse motion relative to the fixed stars background around October, 20 2021.

We were also able to find the maximum angular velocity of the step motors, which resulted in 36.5° per second, which is fast enough to scan large constellations in a short amount of time.

Those tests were repeated at two other dates with different calibration star sets, all of them resulting in similar results. Moreover, after the initial calibration, no loss in accuracy was observed even after two hours of operation.

As a hardware issue, we noticed that the intensity of the laser beam was heavily temperature dependent, which required frequent adjustments to the trim-pot used for current control. Therefore, we plan to use the second microcontroller core available on the ESP32 board to implement a constant current driver for the laser.

Another limitation observed was related to the torsion springs used for tensioning the motor shafts. As they were 3D printed using PLA filament, they showed a change in elasticity when keeping stretched for a long time. A solution to this problem may be their replacement with metal springs.

\section{Conclusions}

With this work we hope that we have showed the potential Stellector has to be used in non formal educational environments by schools, universities, astronomical observatories and science museums. As such, it is a project under continuous development, and it can benefit from the feedback given by people and their interaction with the equipment. Unfortunately, due to the Covid-19 pandemic, we couldn't apply it to the public and further studies will be necessary to fully appreciate its potentials and limitations.

One of the limitations that we can anticipate for a wider use of Stellector is the need to inform the position of a relatively large number of celestial bodies for system calibration (around five stars, as discussed in section \ref{sec:experimental}), which requires some knowledge of the night sky. This difficulty can be overcome by using a printed star map or a night sky web app on a smartphone as a guide. Another possibility is to implement in the Hathor software a routine to identify the geographic location of Horus via GPS and its orientation using the accelerometer and magnetometer contained in the GY-511 sensor at the base of Horus. Thus, it is possible, in principle, for the system to automatically point the laser close to each calibration star and just ask the user to make fine adjustments in the position of the steppers, as is done in GoTo systems.

Other software and hardware improvements are already under development, as the construction of routines for getting information about the region of the sky, or celestial object, to where the user chooses to point the laser, the implementation of a constant current driver for the laser, and the construction of a protective cover -- also 3D printed -- for the Horus hardware.

The Stellector project is being hosted on Github under a public license \cite{stellector2021}, where all its components and updates are publicly available.

\section*{References}

\bibliography{references}{}
\bibliographystyle{iopart-num}

\end{document}